\documentclass[10pt]{iopart}

\usepackage{iopams}  
\usepackage{graphicx}
\usepackage{subfigure}
\usepackage{bm}
\usepackage{dsfont}
\usepackage{xcolor}
\usepackage[colorlinks=true, linkcolor=olive,citecolor=olive]{hyperref}
\begin{document}


\title[Wave propagation in rotating magnetised plasmas]{Wave propagation in rotating magnetised plasmas}

\author{Renaud Gueroult}
\address{LAPLACE, Universit\'{e} de Toulouse, CNRS, INPT, UPS, 31062 Toulouse, France}
\author{Jean-Marcel Rax}
\address{ACEE, Princeton University, Princeton, NJ 08544, USA}
\address{IJCLab, Université de Paris-Saclay, 91405 Orsay, France}
\author{Nathaniel J. Fisch}
\address{Department of Astrophysical Sciences, Princeton University, Princeton, New Jersey 08540, USA}
\ead{renaud.gueroult@laplace.univ-tlse.fr}
\vspace{10pt}
\begin{indented}
\item[]September 2022
\end{indented}

\begin{abstract}
Wave propagation properties in a medium are fundamentally affected when this medium is moving instead of at rest. In isotropic dielectric media rotation has two noteworthy contributions: one is a mechanically induced circular birefringence which materialises as a rotation of the polarisation, the other is image rotation which corresponds to a rotation of the transverse structure of a wave. Here we review the effect of rotation in a magnetised plasma. We first show that the mechanical effect of rotation on polarisation is in a magnetised plasma superimposed onto the classical Faraday rotation, and that failing to account for this new contribution could lead to errors in the interpretation of polarimetry data. We also demonstrate that image rotation is recovered in plasmas for a number of low-frequency magnetised plasma waves carrying orbital angular momentum, and that this phenomenon holds promise for the development of new rotation diagnostic tools in plasmas.

\end{abstract}

%
%
%
%
\ioptwocol

\section{Introduction}

Wave propagation properties in a moving medium differ from that in this same medium at rest. This fundamental result was first postulated by Fresnel~\cite{Fresnel1818} and later demonstrated experimentally by Fizeau~\cite{Fizeau1851} by observing the interference pattern formed by light beams propagating along and against a water flow. As seen from an observer in the laboratory frame in which the medium is moving, the beam propagating along the flow appears to propagate faster than that propagating against the flow, giving the appearance that the light is dragged by the medium. Although Fresnel's hypothesis of aether drag proved to be wrong, the predicted formula for the dragging coefficient held and was later shown by von Laue to be consistent with relativity theory~\cite{vonLaue1907}, and this effect is accordingly still referred to as longitudinal Fresnel drag. 

While Fresnel and Fizeau's original work focused exclusively on the case of a longitudinal linear motion, that is $\mathbf{k}\parallel\mathbf{v}$ with $\mathbf{k}$ the wave vector and $\mathbf{v}$ the medium's constant uniform velocity in the observer's frame, the case of a transverse linear motion, that is $\mathbf{k}\perp\mathbf{v}$, was later considered~\cite{Jones1972}. In this case it is the optical path that is altered, with a group velocity in the laboratory frame that now has a component along the direction of motion of the medium. A beam passing through a medium moving perpendicularly to the wave vector thus exits the medium displaced in the direction of motion, as demonstrated experimentally by Jones~\cite{Jones1972,Jones1975}. By analogy with the longitudinal effect identified by Fresnel, and because the dragging coefficient is the same~\cite{Rogers1975,Player1975}, this effect is referred to as transverse Fresnel drag. 

A situation closely related to transverse Fresnel drag, though different in that the motion is not uniform, is that of a wave propagating along the rotation axis of a rotating medium. Interestingly this configuration happened to have attracted the interest of Thomson~\cite{Thomson1885} and then Fermi~\cite{Fermi1923}  years before the transverse Fresnel drag effect was identified. Thomson and Fermi indeed postulated that a linearly polarised wave propagating along the axis of rotation of a rotating medium should see its plane of polarisation rotate. This result was demonstrated experimentally many years later by Jones who examined light propagating along a rotating glass rod~\cite{Jones1976}. It was at the same time interpreted as a mechanically induced circular birefringence - that is a difference in real refractive wave index for circularly polarised eigenmodes - by Player~\cite{Player1976}. By analogy with longitudinal and transverse drags, this effect is often referred to as polarisation drag.

Because circularly polarised eigenmodes corresponds to the wave's spin angular momentum (SAM) states of $\pm\hbar$, polarisation drag can be understood as a phase shift introduced by rotation between eigenmodes of opposite SAM content. Expanding on this idea, and noting that wave can carry orbital angular momentum (OAM) in addition to SAM~\cite{Allen1992a}, Padgett \emph{et al.} postulated that rotation should similarly introduce a phase shift between eigenmodes of opposite OAM content $\pm l\hbar$, which should materialise as a rotation of the transverse structure of the wave~\cite{Padgett2006}. Generalising Player's derivation to wave's carrying OAM, G{\"o}tte \emph{et al.} showed~\cite{Goette2007} that in the case of an isotropic dielectric with rest-frame wave phase index $n'_{\phi}(\omega)\sim1$ the angle by which the polarisation and the transverse structure are rotated is the same and equal to
\begin{equation}
\Delta\theta = \left(n'_{g}-{n'_{\phi}}^{-1}\right)\frac{\Omega L}{c},
\label{Eq:polarisation_rotation}
\end{equation}
with $n'_{g}$ the rest-frame group index of the medium, $\Omega$ the rotation frequency of the medium, $L$ the propagation length in the medium and $c$ the speed of light. This effect of rotation on wave carrying OAM, known as image rotation, was subsequently demonstrated experimentally by taking advantage of slow-light conditions in a rotating ruby rod~\cite{Franke-Arnold2011}. 

Short of singular conditions though, the polarisation rotation angle $\Delta\theta$ introduced by rotation is small. This is because in conventional isotropic dielectrics $n'_{\phi}(\omega)$ and $n'_{g}$ are typically $\mathcal{O}(1)$, and the angular velocity $\Omega$ of a solid medium is limited by mechanical constraints. Quantitatively, Jones measured polarisation rotations of about 1 micro radian after crossing four times a $10$~cm long glass rod. From (\ref{Eq:polarisation_rotation}), directions to enhance this effect can be divided into two categories, namely media with large group or phase index and higher angular frequency $\Omega$. The former has been exploited in solid media by creating slow light conditions through electromagnetically induced transparency~\cite{Fleischhauer2005,Carusotto2003,Kuan2016,Solomons2020} or coherent population oscillations~\cite{Bigelow2003a,Bigelow2003,Franke-Arnold2011}, which can then lead to effective group index $n'_{g}$ in excess of $10^{6}$. The latter lead to the use of dilute media, notably super-rotors~\cite{Steinitz2020,Milner2021,Tutunnikov2022} that can achieve $\Omega\sim10^{14}$~rad/s.

Plasmas are particularly interesting in this context in that they can in principle take advantage of both of these options at once. Indeed, the dispersive properties of plasmas offer opportunities rotation for larger phase and group index, and rotation in a plasma can be produced electromagnetically, removing the need for moving parts and allowing in turn for high angular velocities. On the other hand, plasmas are gyrotropic media, so that rotation effects on wave's SAM are superimposed on SAM properties in this same gyrotropic medium at rest.

In this paper we summarize what has been learned in recent years about the effect of rotation on wave propagation in magnetised plasmas. Section~\ref{Sec:lab_rot_frame} first recalls some basic elements and definitions on propagation in a magnetised plasma at rest, as well as reviews the two main approaches used to study propagation in moving media. Section~\ref{Sec:SAM} then discusses the effect of rotation on the wave's polarisation, that is to say its spin angular momentum, and points to different environments where these effects could be of importance. Section~\ref{Sec:OAM} discusses the effect of rotation on the wave's orbital angular momentum and considers the opportunities this effect may hold for rotation diagnostic purposes. Finally, section \ref{Sec:summary} summarises the main findings of this study.

\section{Definitions and models}
\label{Sec:lab_rot_frame}

\subsection{Circular birefringence in a magnetised plasma at rest}

Before embarking on the analysis of the effect of rotation, we briefly recall a number of definitions and results on the optical activity of a magnetised plasma at rest.
 
First let us write
\begin{equation}
\hat{\bar{\chi}}(\omega) = \left(\begin{array}{c c c}
\bar{\chi}_{\perp} & -i\bar{\chi}_{\times} & 0\\\
i\bar{\chi}_{\times} & \bar{\chi}_{\perp} & 0\\
0 & 0 & \bar{\chi}_{\parallel}
\end{array}\right)
\label{Eq:dielectric_tensor}
\end{equation}
the susceptibility tensor of a cold magnetised plasma at rest in an inertial frame,  with components
\begin{eqnarray}
\bar{\chi}_{\perp}(\omega)  = \sum\limits_{\alpha}\frac{{\omega_{p\alpha}}^2}{{\Omega_{c\alpha}}^2-\omega^2},\label{Eq:Tensor_component_perp}\\ 
\bar{\chi}_{\times}(\omega)  = \sum\limits_{\alpha}\epsilon_{\alpha}\frac{\Omega_{c\alpha}}{\omega}\frac{{\omega_{p\alpha}}^2}{\omega^2-{\Omega_{c\alpha}}^2},\label{Eq:Tensor_component_cross}\\
\bar{\chi}_{\parallel}(\omega)  = -\sum\limits_{\alpha}\frac{{\omega_{p\alpha}}^2}{\omega^2}\label{Eq:Tensor_component_para},
\end{eqnarray}
where $\Omega_{c\alpha} = |q_\alpha|B_0/m_\alpha$ and $\omega_{p\alpha} = [n_{\alpha} e^2/(m_{\alpha}\epsilon_0)]^{1/2}$ are respectively the unsigned cyclotron frequency and plasma frequency of species $\alpha$, and $\epsilon_{\alpha} = q_\alpha/|q_\alpha|$. Parallel and perpendicular directions are defined here with respect to the background magnetic field $\mathbf{B}_{0} = B_{0}\mathbf{\hat{e}}_{z}$. Seeking plane wave solutions in the form $\mathbf{A} = \bm{A}\exp[i(\mathbf{k}\cdot\mathbf{r}-\omega t)]$, Maxwell-Faraday and Maxwell-Ampère write
\begin{eqnarray}
\mathbf{k}\times\mathbf{E}' &= \omega \mathbf{B}',\\
\mathbf{k}\times\mathbf{H}' &= -\omega \mathbf{D}' ,
\end{eqnarray}
which with the help of the constitutive relations $\mathbf{D}'=\epsilon_{0}(\mathds{1}+\hat{\bm{\bar{\chi}}})\cdot\mathbf{E}'$ and $\mathbf{B}'=\mu_{0}\mathbf{H}' $ can be combined to give
\begin{equation}
\left[\left(\frac{c^{2}k^{2}}{\omega^{2}}-1\right)\mathds{1}-\frac{c^{2}}{\omega^{2}}\mathbf{k}\otimes\mathbf{k} -\hat{\bar{\chi}}\right]\mathbf{E}' = \hat{\bm{\mathcal{D}}}\cdot\mathbf{E}'  = 0
\end{equation}
where we used that $\mathbf{k}\times\mathbf{k}\times~ = -k^{2}\mathds{1}+\mathbf{k}\otimes\mathbf{k}$. Non-trivial solutions demand $det(\hat{\bm{\mathcal{D}}})=0$. Focusing more specifically on parallel propagation $\mathbf{k} = ~^{T}(0,0,k)$, that is a wave vector along the background magnetic field, this condition writes
\begin{equation}
\left|\begin{array}{c c c}
n^{2}-1-\bar{\chi}_{\perp} & i\bar{\chi}_{\times} & 0\\
-i\bar{\chi}_{\times} & n^{2}-1-\bar{\chi}_{\perp} & 0\\
0 & 0 & -1-\bar{\chi}_{\parallel}\\
\end{array}\right| = 0
\end{equation}
with $n = kc/\omega$ the refractive index. The solutions are
\begin{equation}
\bar{n}^{2}(\omega) = 1+ \bar{\chi}_{\perp}(\omega) \pm \bar{\chi}_{\times}(\omega),
\end{equation}
and one verifies that the $\pm$ solutions correspond respectively to right-circularly polarised (RCP) and left-circularly polarised (LCP) eigenmodes $\mathbf{E} = E_{w}(\mathbf{\hat{e}}_{x}\pm i\mathbf{\hat{e}}_{y})\exp[i(\mathbf{k}\cdot\mathbf{r}-\omega t)]$, with left- and right-handed polarised waves defined here from the point of view of the source in the direction of propagation of the wave $\mathbf{k}$. We  write accordingly
\begin{eqnarray}
\mbox{$\bar{n}_{rcp}$}^{2}(\omega) = 1+ \bar{\chi}_{\perp}(\omega) + \bar{\chi}(\omega)_{\times},\label{Eq:index_RCP_non_moving_anisotropic}\\
\mbox{$\bar{n}_{lcp}$}^{2}(\omega) = 1+ \bar{\chi}_{\perp}(\omega) - \bar{\chi}(\omega)_{\times}\label{Eq:index_LCP_non_moving_anisotropic}.
\end{eqnarray}

From (\ref{Eq:index_RCP_non_moving_anisotropic}) and (\ref{Eq:index_LCP_non_moving_anisotropic}) the difference in wave index $\bar{n}_{rcp}$ and $\bar{n}_{lcp}$ of RCP and LCP waves comes from the non zero off-diagonal term $\bar{\chi}_{\times}$. This difference in wave index translates into a difference in phase velocity $v_{\phi} = c/n$, which is the source of a rotation of the plane of polarisation of a linearly polarised wave~\cite{Fresnel1822}. After propagating over a distance $l$ along $\mathbf{k}$, the polarisation has been rotated by an angle
\begin{equation} 
\Delta \phi(\omega)  = \left[\bar{n}_{lcp}(\omega)-\bar{n}_{rcp}(\omega)\right]\frac{\omega l}{2c}.
\label{Eq:general_polarization_rotation}
\end{equation}
The polarisation rotation per unit length, also known as the specific rotary power, then writes 
\begin{equation} 
\delta (\omega) \doteq \frac{\Delta \phi(\omega)}{l} = \left[\bar{n}_{lcp}(\omega)-\bar{n}_{rcp}(\omega)\right]\frac{\omega}{2c}.
\label{Eq:general_rotatory_power}
\end{equation}

Going back to (\ref{Eq:Tensor_component_cross}) we see that, other than for the particular case where the contributions to $\bar{\chi}_{\times}$ of the different species $\alpha$ cancel each others, the background magnetic field $\mathbf{B}_{0}$ is responsible for a non-zero $\bar{\chi}_{\times}$. This underlines the gyrotropic nature of plasmas, and the associated polarisation rotation is the well known Faraday rotation~\cite{Faraday1846}. Faraday rotation in plasmas, especially in astrophysics, is often considered in the limit that $\Omega_{c\alpha}\ll\omega$ and $\omega_{p\alpha}\ll\omega$. In this limit, $1\gg |\bar{\chi}_{\perp}| \gg |\bar{\chi}_{\times}|$, $\bar{\chi}_{\perp}<0$ and $\bar{\chi}_{\times}<0$, so that $n_{lcp}(\omega)\geq n_{rcp}(\omega)$ when $B_0>0$ and, from (\ref{Eq:general_polarization_rotation}), $\Delta \phi>0$. 
Quantitatively, 
\begin{equation}
\bar{n}_{lcp}(\omega)- \bar{n}_{rcp}(\omega)\sim \frac{\Omega_{ce}{\omega_{pe}}^2}{\omega^3},
\label{Eq:HF_Faraday}
\end{equation}
which from (\ref{Eq:general_polarization_rotation}) yields the classical scaling $\Delta \phi \propto \lambda^2$ with $\lambda = 2\pi/k$ the wavelength. Note though that this simple scaling is only valid in the limit that the wave frequency is much larger than the plasma natural frequencies $\Omega_{ce}$ and $\omega_{pe}$. 

\subsection{Laboratory vs. rotating frame modelling}

Having briefly reviewed propagation properties in a plasma at rest, we now turn to the effect of motion. We note respectively $\Sigma$ and $\Sigma'$ the laboratory frame and the plasma rest-frame, and use a prime notation to refer to variables expressed in $\Sigma'$. Previous work aiming at determining the electromagnetic properties of a moving plasma can be divided into two groups according to which frame of reference ($\Sigma$ or $\Sigma'$ ) is used. 

A first group gathers studies examining propagation in the plasma rest-frame or rotating frame $\Sigma'$. This is historically the first method used. In the particular case of rotation, this method was notably used by Lehnert using MHD~\cite{Lehnert1954,Lehnert1955} and then by Tandon and Bajaj using a two fluid model and Maxwell's equations~\cite{Tandon1966}. The two fluid model was then extended by Uberoi and Das~\cite{Uberoi1970} to account for collisional effects, and by Verheest to include thermal effects in the form of a scalar pressure~\cite{Verheest1974}.  In all these studies though rotation was only modelled through the Coriolis force. This choice appears to be guided by Chandrasekhar's proposition that the Coriolis force may play a predominant role in astrophysics~\cite{Chandrasekhar1953}, in contrast to laboratory scale phenomena where it is usually negligible. This however clearly questions the validity of the obtained results in regimes where centrifugal effects are expected to be important. In addition, Engels and Verheest~\cite{Engels1975} argued that the use of the standard form of Maxwell's equations in these studies was incorrect, and that one should instead use Schiff's current and charge densities~\cite{Schiff1939,Webster1963}. A basic issue at the root of this problem is that the covariant equations of electrodynamics do not uniquely determine the three-vector formulation of the electromagnetic equations in a (non-inertial) rotating frame of reference~\cite{Gron1984}. As a result, there remains a question of which systems of equations hold in the plasma rest-frame.

A second group gathers studies examining propagation in the laboratory frame $\Sigma$, with two different approaches. A first approach is to consider the transformation of the various parameters from the rotating plasma rest-frame to the laboratory frame. This is the approach originally proposed by Player~\cite{Player1976}. A challenge here though is that one has to decide what is the medium's response in its rest-frame, that is using our notations how  $\hat{\chi}'(\omega')$ relates to $\hat{\bar{\chi}}(\omega')$. An assumption that if most often made~\cite{Player1976,Goette2007,Gueroult2019a,Gueroult2020} is that the properties of the medium are unaffected by the medium's motion, that is $\hat{\chi}'(\omega')=\hat{\bar{\chi}}(\omega')$. However Baranova and Zel'dovich postulated that there should actually be an effect of the motion on the refractive index through the Coriolis force~\cite{Baranova1979}, and Nienhuis \emph{et al.} showed that there should be an effect of the motion on the refractive index for a dispersive medium near to an absorption resonance~\cite{Nienhuis1992}. Jones' experimental observations~\cite{Jones1976}, however, were found in agreement with Player's predictions~\cite{Player1976}. A second approach consists in determining from first principles the dynamics of charged particles in the laboratory frame, and to then use this equilibrium to expose the linear response of this moving plasma. In the particular case of rotation this is the method we recently used to study orbital angular momentum waves in a rotating plasma~\cite{Rax2021}. 

In this paper we focus on the latter, that is we work in the laboratory frame $\Sigma$. Lorentz transforms of fields from the rotating plasma rest-frame to the laboratory frame are used to expose the effect of rotation on the wave's spin angular momentum in Section~\ref{Sec:SAM}, whereas the linear response of a moving plasma computed in the laboratory frame is used to expose the effect of rotation on the wave's orbital angular momentum in Section~\ref{Sec:OAM}.

\section{Spin angular momentum effects in a rotating magnetised plasma}
\label{Sec:SAM}

\subsection{Circular birefringence in a rotating magnetised plasma}

\begin{figure}
\begin{center}
\includegraphics[]{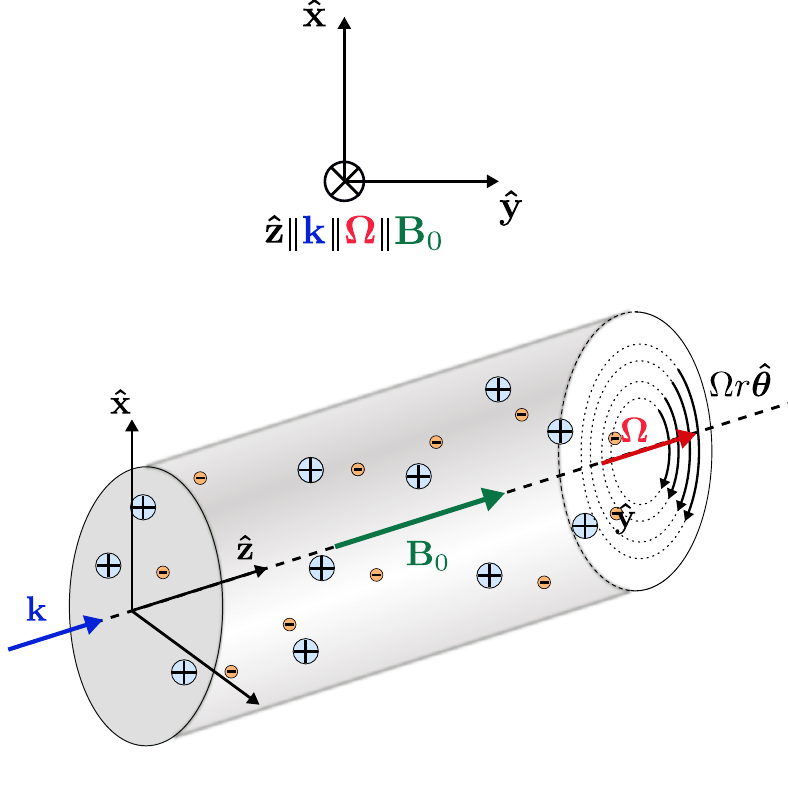}
\caption{Rotating magnetised plasma for which the rotation axis $\bm{\Omega}$, the background magnetic field $\mathbf{B}_{0}$ and the wave vector $\mathbf{k}$ are aligned. This configuration is referred to as an aligned rotator. }
\label{Fig:AlignedRotator}
\end{center}
\end{figure}

As mentioned above, one strategy to derive wave propagation properties in a rotating medium is to start from Maxwell's equation combined with the constitutive equations
\begin{eqnarray}
\mathbf{D'} & = \mathbf{D'} (\mathbf{E'} ,\mathbf{B'})\label{Eq:constitutive_D}\\
\mathbf{H'} & = \mathbf{H'} (\mathbf{E'} ,\mathbf{B'})\label{Eq:constitutive_H}
\end{eqnarray}
relating the field quantities in the medium's rest frame, and then to use an instantaneous Lorentz transform from a local comoving inertial frame to the laboratory frame to obtain the wave equations. This approach is precisely the one used used by Player~\cite{Player1976} and G{\"o}tte~\cite{Goette2007} \emph{et al.} considering an isotropic dielectric with scalar relative permittivity $\epsilon=1+\chi$. In the case of a magnetised plasma, the situation is more intricated in that as shown in (\ref{Eq:dielectric_tensor}) the susceptibility is no longer a scalar $\chi$ but instead a non-diagonal tensor $\hat{\bar{\chi}}$. Yet, it has been shown recently~\cite{Gueroult2019a,Gueroult2020} that this same method can be used to obtain a dispersion relation for a transverse wave propagating along the rotation axis of a magnetised plasma such that $\bm{\Omega}/\Omega = \mathbf{k}/k = \mathbf{B}_{0}/B_{0} = \mathbf{\hat{e}}_{z}$. This configuration, shown in figure~\ref{Fig:AlignedRotator}, is referred to as an aligned rotator. In this case it was found that the eigenmodes are left- and right-circularly polarised, with phase index
\begin{eqnarray}
{n_{rcp}}^2(\omega)  & = & 1 + \bar{\chi}_{\perp}(\omega-\Omega) + \bar{\chi}_{\times}(\omega-\Omega) \label{Eq:index_general_r}\\
 & & -\frac{\Omega}{\omega}\left[\bar{\chi}_{\times}(\omega-\Omega) + \bar{\chi}_{\parallel}(\omega-\Omega)+\bar{\chi}_{\perp}(\omega-\Omega)\right],\nonumber\\
{n_{lcp}}^2(\omega)  & = & 1 + \bar{\chi}_{\perp}(\omega+\Omega) - \bar{\chi}_{\times}(\omega+\Omega) \label{Eq:index_general_l}\\
 & & -\frac{\Omega}{\omega}\left[\bar{\chi}_{\times}(\omega+\Omega) - \bar{\chi}_{\parallel}(\omega+\Omega)-\bar{\chi}_{\perp}(\omega+\Omega)\right].\nonumber
\end{eqnarray}
We stress here again that a key assumption in deriving these results, just like those previously obtained by Player~\cite{Player1976} and G{\"o}tte~\cite{Goette2007},  is that the properties of the medium are unaffected by the rotation (\emph{i.~e.} $\hat{\chi}'(\omega')=\hat{\bar{\chi}}(\omega')$). 

Comparing (\ref{Eq:index_RCP_non_moving_anisotropic}) and (\ref{Eq:index_LCP_non_moving_anisotropic}), that is the wave indexes absent rotation, with respectively (\ref{Eq:index_general_r}) and (\ref{Eq:index_general_l}), we see that rotation has two effects. One is to add a supplemental contribution to the wave index. This is the last term in brackets on the right hand side of  (\ref{Eq:index_general_r}) and (\ref{Eq:index_general_l}). Because this extra contribution depends on the handedness, and since we have showed that circular birefringence arises from a difference in phase index between left- and right- circularly polarised eigenmodes, we conclude that mechanical rotation will affect polarisation rotation. This general result will be discussed in more details through various simplifying assumptions in the next paragraphs. The other effect of rotation on wave indexes is that the frequency that appears in the dielectric response is no longer the laboratory frame frequency $\omega$ but instead the angular Doppler shifted frequency $\omega' = \omega\mp\Omega$ that is measured in the plasma rest-frame~\cite{Garetz1979,Garetz1981}. This is expected to become particularly important for rotation frequencies that are comparable with the wave frequency. Finally, we verify as expected that (\ref{Eq:index_general_r}) and (\ref{Eq:index_general_l}) reduce to (\ref{Eq:index_RCP_non_moving_anisotropic}) and (\ref{Eq:index_LCP_non_moving_anisotropic}) in the limit $\Omega=0$.

Trying to make sense of (\ref{Eq:index_general_r}) and (\ref{Eq:index_general_l}), one can use the fact that for low-frequency waves $\omega\ll\Omega_{ci}$ and slow rotation $\Omega\ll\omega$ the components of the susceptibility tensor write to leading order in $\omega/\Omega_{ci}\ll1$
\begin{eqnarray}
\bar{\chi}_{\perp}(\omega\mp\Omega)  = \frac{{\omega_{pi}}^{2}}{{\Omega_{ci}}^{2}},\label{Eq:Tensor_component_perp_low}\\ 
\bar{\chi}_{\times}(\omega\mp\Omega)  = -\frac{{\omega_{pi}}^{2}}{{\Omega_{ci}}^{2}}\frac{\omega}{\Omega_{ci}},\label{Eq:Tensor_component_cross_low}\\
\bar{\chi}_{\parallel}(\omega\mp\Omega)  = -\frac{1}{\eta^{2}}\frac{{\omega_{pi}}^{2}}{{\Omega_{ci}}^{2}}\frac{{\Omega_{ci}}^{2}}{\omega^{2}}\label{Eq:Tensor_component_para_low},
\end{eqnarray}
with $\eta^{2}\ll1$ the electron to ion mass ratio. This implies that  (\ref{Eq:index_general_r}) and (\ref{Eq:index_general_l}) reduce in this frequency band and for $\Omega\ll\omega$ to
\begin{eqnarray}
{n_{rcp}}^2(\omega)  & = \mbox{$\bar{n}_{rcp}$}^{2}(\omega)-\frac{\Omega}{\omega} \bar{\chi}_{\parallel}(\omega),\label{Eq:index_low_r}\\
{n_{lcp}}^2(\omega)  & = \mbox{$\bar{n}_{lcp}$}^{2}(\omega)+\frac{\Omega}{\omega} \bar{\chi}_{\parallel}(\omega).\label{Eq:index_low_l}
\end{eqnarray}

\begin{figure}
\begin{center}
\includegraphics[]{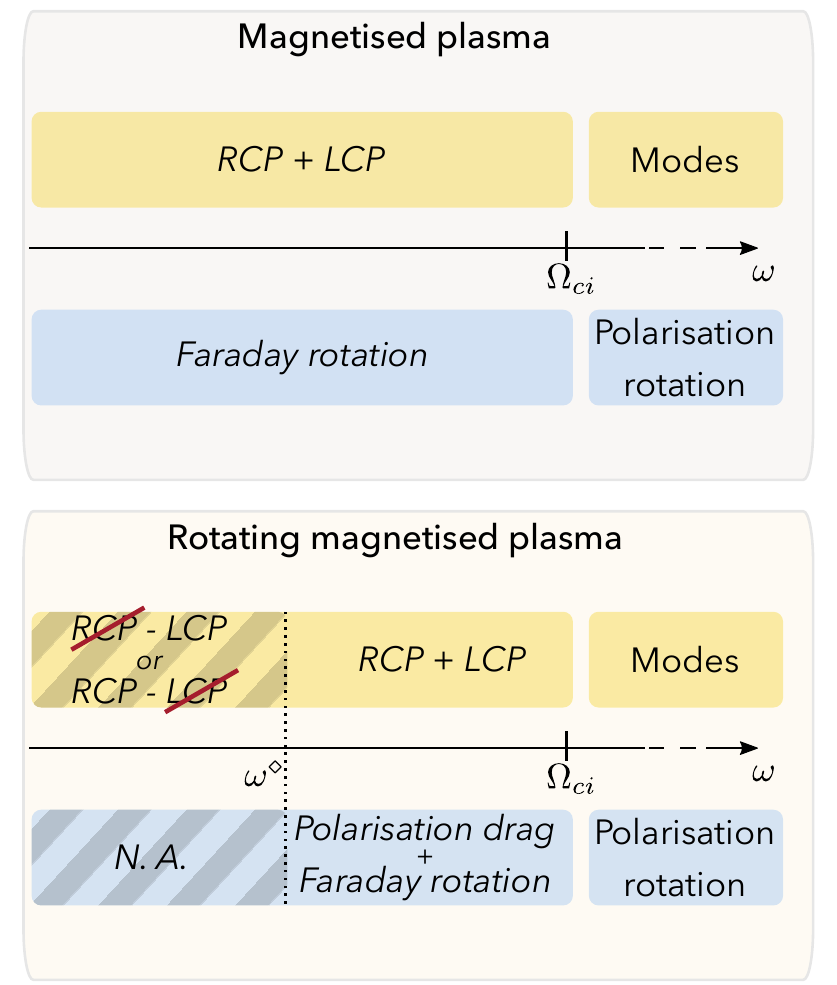}
\caption{Sketch of the low-frequency $\omega\ll\Omega_{ci}$ behaviour of a slowly rotating $\Omega\ll\omega$ aligned rotator (bottom), as compared to the classical behaviour of a magnetised plasma at rest (top). Rotation introduces a low-frequency cutoff $\omega^{\diamond}$ below which one of the two circularly polarised eigenmodes does not propagate. Above the cutoff polarisation rotation is the sum of Faraday rotation and polarisation drag. }
\label{Fig:PolarisationDrag}
\end{center}
\end{figure}

Examining the terms in (\ref{Eq:index_low_r}) and (\ref{Eq:index_low_l}), the rest-frame wave index \mbox{$\bar{n}_{rcp/lcp}$} tends to a constant equal to $(1+{\omega_{pi}}^{2}/{\Omega_{ci}}^{2})^{1/2}$ at low frequency, whereas as shown in (\ref{Eq:Tensor_component_para}) $\bar{\chi}_{\parallel}(\omega)\propto \omega^{-2}$.  This allows to conclude that there is a cutoff at frequency $\omega^{\diamond}$ below which one of the circularly polarised eigenmodes does not propagate (LCP and  RCP for respectively $\Omega>0$ and $\Omega<0$). In the remaining of this study we consider only $\Omega>0$ as the generalisation to $\Omega>0$ is straighforward, and thus a mechanically induced cutoff for the LCP wave. From (\ref{Eq:index_low_l}) the cutoff frequency is solution of
\begin{equation}
\frac{{\omega^{\diamond}}^{4}}{{\Omega_{ci}}^{4}}+\frac{{\omega^{\diamond}}^{3}}{{\Omega_{ci}}^{3}}\left(1+ \frac{{\Omega_{ci}}^{2}}{{\omega_{pi}}^{2}}\right)-\eta^{-2}\frac{\Omega}{\Omega_{ci}} = 0.
\end{equation}
Above this cutoff both mode propagate and the difference in phase indexes $n_{lcp}-n_{rcp}$,
which we see from (\ref{Eq:index_low_r}) and (\ref{Eq:index_low_l}) comes both from the intrinsic gyrotropy $\mbox{$\bar{n}_{lcp}$}(\omega)\neq\mbox{$\bar{n}_{rcp}$}(\omega)$ and from rotation $\Omega\neq0$, is then the source of polarisation rotation. This is illustrated in figure~\ref{Fig:PolarisationDrag}.

\subsection{Under-dense plasmas}

To better expose the effect of rotation, we now focus on under-dense plasmas, and more specifically assume ${\Omega_{ci}}^{2}/{\omega_{pi}}^{2}={v_{A}}^{2}/c^{2}\gg1$, with $v_{A}$ and $c$ respectively the Alfv{\'e}n velocity and the speed of light. The cutoff then reduces to lowest order to 
\begin{equation}
\omega^{\diamond} = \left(\Omega{\omega_{pe}}^{2}\right)^{1/3}.
\label{Eq:under_dense_cutoff}
\end{equation}
From (\ref{Eq:Tensor_component_perp_low}) and (\ref{Eq:Tensor_component_cross_low})  we also have $|\bar{\chi}_{\times}(\omega\ll\Omega_{ci})|\ll\bar{\chi}_{\perp}(\omega\ll\Omega_{ci})\ll1$. For rotation slow enough that $\Omega|\bar{\chi}_{\parallel}(\omega)|/\omega\ll1$ a Taylor expansion of the wave index difference then gives 
\begin{equation}
n_{lcp}-n_{rcp} \sim   \frac{{\omega_{pi}}^{2}}{{\Omega_{ci}}^{2}}\left[\frac{\omega}{\Omega_{ci}}-\frac{{\Omega_{ci}}^{2}}{\omega^{2}}\frac{\Omega}{\omega}\right],
\label{Eq:index_diff_low}
\end{equation}
and, from (\ref{Eq:general_rotatory_power}), the specific rotary power
\begin{equation}
\delta\sim   \frac{{\omega_{pi}}^{2}}{{2\Omega_{ci}}c}\left[\frac{\omega^{2}}{{\Omega_{ci}}^{2}}-\frac{\Omega{\Omega_{ci}}}{\omega^{2}}\right].
\label{Eq:rot_power_low}
\end{equation}
It should however be noted that (\ref{Eq:index_diff_low}) and (\ref{Eq:rot_power_low}) are only valid for $\omega\gg\omega^{\diamond}$ since the condition $\Omega|\bar{\chi}_{\parallel}(\omega)|/\omega\ll1$ is not verified at the cutoff. 

The first term in brackets in (\ref{Eq:index_diff_low}) comes from the intrinsic gyrotropy $\bar{\chi}_{\times}$.  Note that the apparent inverse scaling on the magnetic field seen here through ${\Omega_{ci}}^{-3}$  is only the consequence of the Taylor expansion of $\bar{\chi}_{\times}$ at low frequency. This term is the source of Faraday rotation. The  second term in brackets in (\ref{Eq:index_diff_low}) comes from the plasma rotation, as underlined by its dependence on $\Omega$. This is the source of polarisation drag. Note here that interestingly, just like as shown through (\ref{Eq:HF_Faraday}) high-frequency Faraday rotation scales as the wavelength square, high frequency polarisation drag scales as the wavelength square. 

From (\ref{Eq:index_diff_low}) one also finds that there is a frequency 
\begin{equation}
\omega^{\star}  = \left(\Omega{\Omega_{ci}}^{3}\right)^{1/4}
\label{Eq:transition}
\end{equation}
for which the contributions to polarisation rotation are equal and cancel out. For $\omega^{\diamond}\leq\omega\leq\omega^{\star}$ polarisation drag dominates over Faraday rotation. Conversely, Faraday rotation dominates over polarisation drag for $\omega\geq\omega^{\star}$. Comparing $\omega^{\diamond}$ and $\omega^{\star}$, one finds 
\begin{equation}
\frac{\omega^{\star}}{\omega_{\diamond}}  = \eta^{2/3}\left(\frac{\omega_{pe}}{\Omega}\right)^{1/12}\left(\frac{\Omega_{ci}}{\omega_{pi}}\right)^{3/4}
\end{equation}
which confirms that there exists a frequency band where polarisation drag is dominant in the limit of a slowly rotating underdense plasma considered here.

Lastly, it is possible to Taylor expand the wave indexes for small $(\omega-\omega^{\diamond})/\omega^{\diamond}$, that is near the cutoff where the $n_{l}=0$. In this case one finds 
\begin{equation}
n_{lcp}-n_{rcp} \sim  -\sqrt{2}+\sqrt{3\frac{\omega-\omega^{\diamond}}{\omega^{\diamond}}},
\label{Eq:index_diff_cut_off}
\end{equation}
revealing that polarisation drag near the cutoff exhibits a different scaling than the wavelength square dependence found to hold at higher frequency.

This behaviour across the low wave frequency range $[\omega_{\diamond},\omega^{\star}]$ is verified when plotting the specific rotatory power computed from the full wave indexes (\ref{Eq:index_general_r}) and (\ref{Eq:index_general_l}), as shown in figure~\ref{Fig:PolarisationDragPlot}. We verify that there is indeed a low frequency region just above the cutoff $\omega_{\diamond}$ where polarisation drag is the dominant contribution to polarisation rotation. This behaviour persists up until about $\omega^{\star}$, at which point polarisation drag becomes negligible in front of Faraday rotation.  The fact that the transition from polarisation drag dominated to Faraday rotation dominated is not observed to occur exactly for $\omega=\omega^{\star}$ in figure~\ref{Fig:PolarisationDragPlot} comes from the fact that (\ref{Eq:transition}) is derived from (\ref{Eq:index_diff_low}), which as mentioned above is only valid in the limit that $\omega\gg\omega^{\diamond}$. Here this ratio is only about $10$.

\begin{figure}
\begin{center}
\includegraphics[]{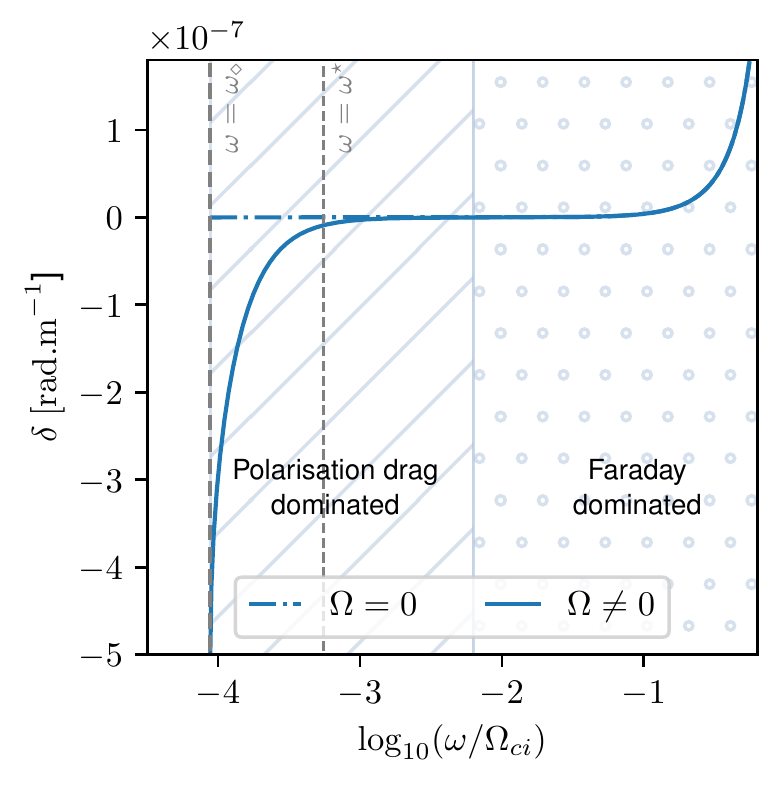}
\caption{Specific rotary power $\delta$ from (\ref{Eq:general_rotatory_power}) as a function of the wave angular frequency $\omega$ for low frequency waves  ($\omega\ll\Omega_{ci}$) in an under-dense plasma, with (solid line) and without (dash-dot line) rotation.  Here we choose $\Omega_{ci}=100\omega_{pi}$ and $\omega^{\star}/\omega_{\diamond}=10$. At low frequency but above the cutoff  $\omega_{\diamond}$ polarisation drag is found to dominate over Faraday rotation. Below the cutoff one of the two circularly polarised eigenmodes does not propagate and as a result $\delta$ is undefined. }
\label{Fig:PolarisationDragPlot}
\end{center}
\end{figure}

\subsection{Implications}

Although the study of polarisation drag effects in plasmas is only at its inception, it is already possible to point to a number of possible implications of the theoretical findings presented above. 

A particularly compelling reason to study these effects concerns pulsars~\cite{Becker2009}, and more particularly pulsar polarimetry.  Pulsars’ highly polarised emission makes them unmatched sources to probe the magnetic fields through Faraday rotation, and pulsar polarimetry has as a result become a standard tool to study magnetic field properties in the interstellar medium (ISM)~\cite{Han2006}. To do so though, one must in principle track the evolution of the polarisation of a pulsar's signal from its source in the magnetosphere to an observer on Earth. Because pulsars' magnetosphere are supposed to be co-rotating with the neutron star, one should in light of the discussion of polarisation drag above account for this effect in the rotating magnetosphere, which has been neglected to date. Exploring this lead, it has been showed recently~\cite{Gueroult2019a} that this new contribution may not be negligible and, importantly, that due to the identical wavelength square scaling of both Faraday rotation and polarisation drag at high frequency this new contribution may be mistakenly attributed to Faraday rotation in the interstellar medium. Capturing and modelling this effect is thus essential to to correct for possible systematic errors in interstellar magnetic field estimates. Furthermore, disambiguating these two contributions could offer a unique means to determine the rotation direction of pulsars. One promising option here will be to use the different frequency scaling of polarisation drag near the cutoff shown through (\ref{Eq:index_diff_cut_off}). A more detailed discussion of polarisation drag in the context of pulsars can be found in~\cite{Gueroult2019a}. 

Another interesting prospect for waves in rotating plasmas is to combine rotation with the unique properties of plasmas to enable new ways to manipulate light. The ability to control light polarisation through nonreciprocal elements is indeed key to many applications, for instance in the form of optical isolators in telecommunication systems. Materials and physical processes classically used to achieve this control however often set limits on the range of applicability of these technologies. For instance, nonreciprocal properties of ferrites are hardly tuneable, and losses limit their applicability to GHz frequencies and below. As mentioned in introduction, plasmas on the other hand have the conceptual advantage of enabling broadband operation and having limited losses. Following this idea, it has been shown recently that the enhanced polarisation drag effect observed in a rotating plasma above the cutoff could under certain conditions lead to unprecedented nonreciprocal properties in the THz regime~\cite{Gueroult2020}. Similarly to the enhancement found using slow light in a ruby window~\cite{Franke-Arnold2011}, the source of this circular birefringence enhancement in a rotating plasma near the cutoff frequency can be framed as the effect of a very large effective group index. Indeed, the group velocity $d\omega/dk$ of the LCP wave tends to zero as $\omega$ approaches $\omega_{\diamond}$. Beyond shear performances, a key advantage of a rotating plasma based nonreciprocal system is that it could be highly modular. Indeed, the polarisation drag enhancement occurs near the cutoff, which from (\ref{Eq:under_dense_cutoff}) can be controlled both though plasma parameters and rotation. A more detailed discussion of polarisation drag for light manipulation applications can be found in~\cite{Gueroult2020}.

Lastly, it stands to reason that accounting for motion and in particular rotation effects on propagation will prove important for a variety of applications beyond astrophysics and light manipulation. Information on the Doppler shift and the wave index in a complex moving plasma is for instance essential to ensure accurate interpretation of Doppler back-scattering reflectometry data in fusion experiments~\cite{Silva2006,Heuraux2012}.

\section{Orbital angular momentum effects in a rotating magnetised plasma}
\label{Sec:OAM}

\subsection{OAM carrying plasma waves}

Plasma waves carrying orbital angular momentum (OAM) have been a research topic of growing interest in the last decade. In unmagnetised plasmas, OAM-carrying plasma waves have for the most part been studied theoretically under the scalar paraxial approximation~\cite{Mendonca2009,Mendonca2012,Mendonca2012a}, although exact solutions of the vector Maxwell equations have recently been exposed~\cite{Chen2017,Nobahar2019}. For cold magnetised plasmas waves carrying orbital angular momentum have been reported in the form of Trivelpiece-Gould (TG) and Whistler-Helicon (WH) waves~\cite{Stenzel2015,Urrutia2016,Stenzel2016a}, as well as predicted for twisted shear Alfv{\'e}n waves~\cite{Shukla2012}. TG modes have also been observed experimentally in non-neutral plasmas~\cite{Hollmann2000}. Here, within our objective of exposing the effect of rotation in a magnetised plasma, we focus on magnetised plasma waves and consider as an example Whistler-Helicon (WH) waves.

\begin{figure}
\begin{center}
\subfigure[Magnetic perturbation $B_{z}^{+1}$ for $l=1$]{\includegraphics[width=6.5cm]{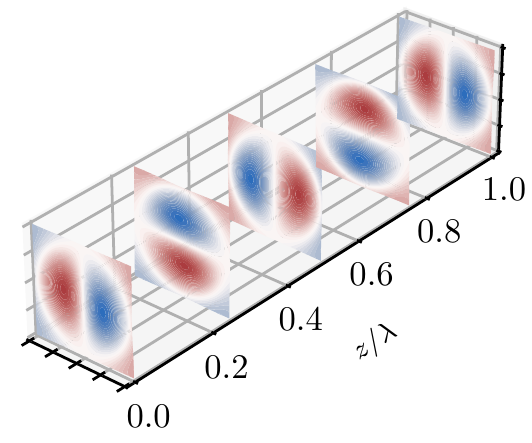}\label{Fig:OAMR}}\\
\subfigure[Magnetic perturbation $B_{z}^{-1}$ for $l=-1$]{\includegraphics[width=6.5cm]{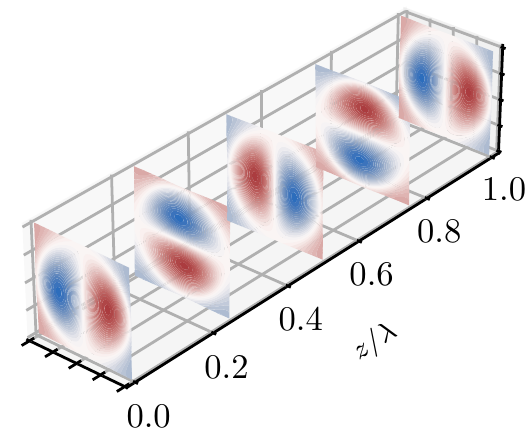}\label{Fig:OAML}}\\
\subfigure[Magnetic perturbation $B_{z}^{+1}+B_{z}^{-1}$ for the sum of $l=1$ and $l=-1$]{\includegraphics[width=6.5cm]{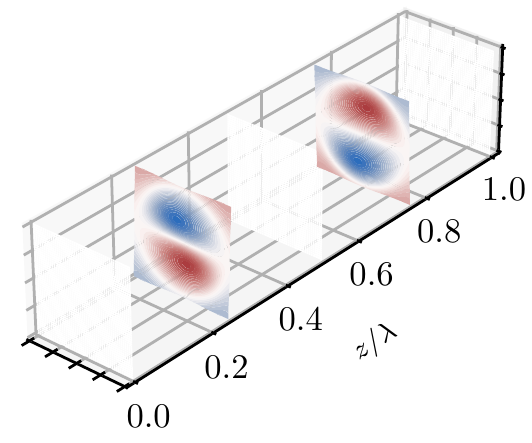}\label{Fig:OAMRL}}
\caption{Transverse structure at a various locations along the axial wave vector $\beta\mathbf{\hat{e}}_{z}$ of \subref{Fig:OAMR} an eigenmode $l=+1$, \subref{Fig:OAML} an eigenmode $l=-1$ and \subref{Fig:OAMRL} the sum of two eigenmodes $l=\pm1$ with equal amplitude. The sign of $l$ dictates the rotation direction of the transverse structure of the wave. The sum of eigenmodes with opposite OAM and equal amplitude is an azimuthally standing wave with zero OAM. The color code simply represents the normalised intensity. $\lambda=2\pi/\beta$ is the axial wavelength. }
\label{Fig:PolarisationDragPlot}
\end{center}
\end{figure}

Waves carrying OAM classically take the form of waves with helical wavefronts  $\exp j(l\theta+\beta z)$ with $l\in\mathbb{Z}$ and $\beta\in\mathbb{R}$, which corresponds to an OAM content $\hbar l$. For instance the axial magnetic perturbation $B_{z}$ associated with WH eigenmodes in a magnetised plasma with $\mathbf{B}=B_{0}\bm{\hat{e}}_{z}$ writes~\cite{Urrutia2016,Rax2021}
\begin{equation}
B_{z}^{l} = J_l(\alpha r)\exp i(l\theta+\beta z-\omega t)
\label{Eq:eigen_l}
\end{equation}
where $J_{l}$ is the ordinary Bessel function of the first kind and of order,  $\alpha$ and $\beta$ are the  radial and axial wave vectors and  $l$ is the azimuthal mode number characterising this eigenmode. For a given mode number $l$ and wave frequency $\omega$ the radial and axial wave vectors  $\alpha$ and $\beta$ depend on one another and on plasma properties via the dispersion relation~\cite{Klozenberg1965,Rax2021}. A characteristic of these modes is that their transverse structure rotates as one moves along the axial wave vector $\beta\mathbf{\hat{e}}_{z}$. Specifically, the direction of rotation depends on the sign of $l$ and thus on the sign of the OAM content of the wave. This can be seen in figures~\ref{Fig:OAMR} and \ref{Fig:OAML} where the eigenmodes $l=+1$ and $l=-1$ are respectively plotted.

If the radial wave structure $\alpha$ is assumed to be set by the antenna, the dispersion relation then determines $\beta$. Interestingly, one finds that in a plasma at rest $\beta$ is independent of $l$~\cite{Klozenberg1965,Rax2021}. A consequence of this result is the following. Consider the sum of eigenmodes of equal amplitude, identical transverse structure $\alpha$ but equal and opposite OAM content $\pm l$. Since $J_{-l}(x) = (-1)^{l}J_{l}(x)$, one gets from (\ref{Eq:eigen_l})
\begin{equation}
B_{z}^{l}+B_{z}^{-l} =  \left\{\begin{array}{l l l}
2iJ_l(\alpha r)\sin(l\theta)e^{ j(\beta z-\omega t)}  &\textrm{if} & l~\textrm{is odd}\\
2J_l(\alpha r)\cos(l\theta)e^{j(\beta z-\omega t)}  &\textrm{if} & l~\textrm{is even}.\\
\end{array}\right.
\label{Eq:sum_modes_norot}
\end{equation}
The combination of these $\pm l$ modes is therefore an azimuthally standing waves~\cite{Stenzel2015a}, which has itself zero OAM. This is illustrated in figure~\ref{Fig:OAMRL}. Drawing an analogy with SAM, we know that the sum of circularly polarised eigenmodes which correspond to equal but opposite spin angular momenta $\pm\hbar$ leads to a linearly polarised wave with zero angular momentum. We now see that the sum of eigenmodes with equal but opposite orbital angular momenta $\pm l\hbar$ leads to a wave with zero angular momentum.

\subsection{Image rotation in a rotating plasma}

\begin{figure}
\begin{center}
\subfigure[$\Omega=0\rightarrow\beta_{-l}=\beta_{+l}$]{\includegraphics[width=8cm]{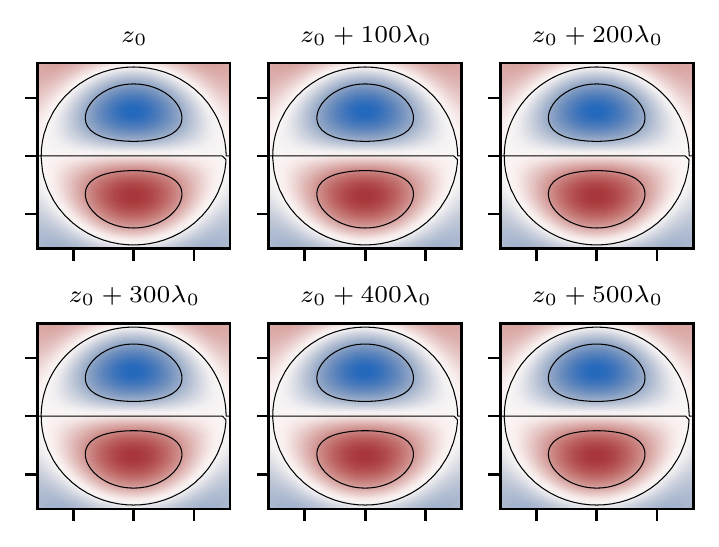}\label{Fig:Image_Rotation_Omega_0}}
\subfigure[$\Omega\neq0\rightarrow\beta_{-l}\neq\beta_{+l}$]{\includegraphics[width=8cm]{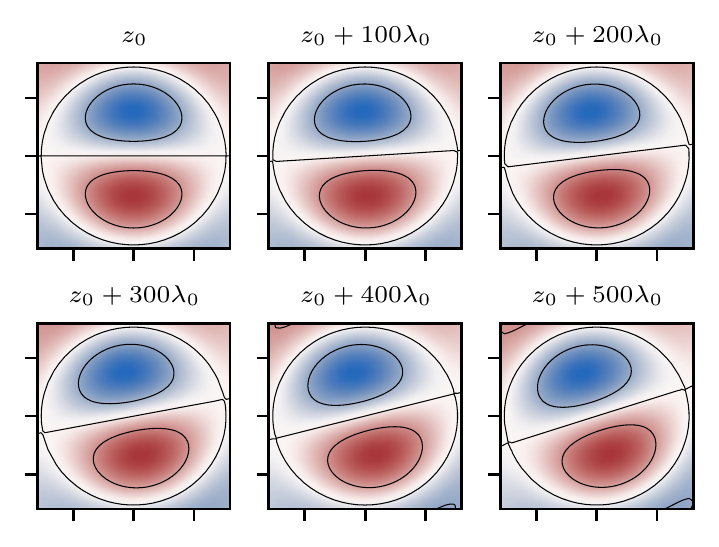}\label{Fig:Image_Rotation_Omega_Neq0}}
\caption{Transverse structure at a various locations along the axial wave vector $\beta\mathbf{\hat{e}}_{z}$ of the sum of two eigenmodes $l=\pm1$ with equal amplitude for \subref{Fig:Image_Rotation_Omega_0} a plasma at rest and \subref{Fig:Image_Rotation_Omega_Neq0} a rotating plasma. In the case of a rotating plasma the transverse structure is seen to rotate as a result of a difference in axial wave vectors $\beta_{-l}\neq\beta_{+l}$ highlighted in (\ref{Eq:axial_wavevector_rot}). This is known as image rotation. Here $\lambda_{0} = 4\pi/(\beta_{-l}+\beta_{+l})$ is the average axial wavelength. }
\label{Fig:ImageRotation}
\end{center}
\end{figure}

In contrast with the case of a plasma at rest for which as mentioned above $\beta$ is independent of $l$, it was shown recently that the propagation of this same wave but in a rotating plasma where $\bm{\Omega}||\mathbf{B}_{0}$ is characterised by an axial wave vector $\beta$ that depends on $l$~\cite{Rax2021}. In particular, it was shown that 
\begin{equation}
\beta(\alpha,l,\Omega,\omega)\neq\beta(\alpha,-l,\Omega,\omega).
\label{Eq:axial_wavevector_rot}
\end{equation} 
If we now go back to our superposition of eigenmodes of equal amplitude, identical transverse structure $\alpha$ but equal and opposite OAM content $\pm l$, we see that as opposed to (\ref{Eq:sum_modes_norot}) the phase of the resulting wave will depend on $\theta$. This is confirmed when plotting the transverse structure of this wave at various locations along along the axial wave vector $\beta\mathbf{\hat{e}}_{z}$, as shown in figure~\ref{Fig:ImageRotation}. The transverse structure of a wave that is azimuthally standing absent rotation, as shown in figure \ref{Fig:Image_Rotation_Omega_0}, is then rotated in the presence of rotation as can be seen in figure \ref{Fig:Image_Rotation_Omega_Neq0}. This is the materialisation in plasmas of the image rotation phenomenon known in dielectrics~\cite{Padgett2006,Franke-Arnold2011}. Due to its analogy both with Faraday rotation and Fresnel drag this effect has been equivalently referred to as mechanical Faraday effect for OAM-carrying beams, rotatory photon drag~\cite{Franke-Arnold2011} or Faraday-Fresnel rotation~\cite{Rax2021}.

This result allows to extend the analogy between SAM and OAM in rotating plasmas further. We have seen in Section~\ref{Sec:SAM} that rotation induces a phase shift between eigenmodes with opposite SAM components (the circularly polarised eigenmodes), which is the source of polarisation rotation. This is polarisation drag. The superposition of two waves with opposite SAM components, which absent rotation would have zero SAM, then has a non-zero SAM when rotation is present. Similarly, we just showed that rotation induces a phase shift between eigenmodes with opposite OAM components, which is the source of a rotation of the transverse structure of a wave. This is image rotation. The superposition of two waves with opposite OAM components, which absent rotation would have zero OAM, then has a non-zero OAM when rotation is present. 

\subsection{Implications}

In contrast with polarisation drag which competes with or supplements Faraday rotation in rotating magnetised plasmas,  an interesting property of image rotation is that it comes exclusively from rotation. This makes the use of OAM carrying waves advantageous compared to SAM carrying waves for plasma rotation diagnostics, as there would be no need to disambiguate which part of rotation comes from rotation and which part comes from the magnetic field alone. Image rotation may then for instance prove useful to diagnose rotation in tokamaks, as it has already been suggested for ultracold atomic gases~\cite{Ruseckas2007} and solid objects~\cite{Lavery2013,Wang2021}.

Although this difference gives a clear conceptual advantage of image rotation over polarisation drag and should motivate further studies, using OAM carrying waves also brings new challenges. First, since image rotation involves a rotation of the wave's transverse structure, the plasma properties along the beam path would have to allow preserving this structure. Although what this exactly implies will likely depend on the particular wave being considered, it stands to reason that some degree of uniformity and structure would be needed on the scale of the beam transverse size. Second, while image rotation indeed comes entirely from rotation and is zero absent rotation, the amount by which the transverse structure is rotated may depend on the magnetic field in the case of a magnetised plasma. This is for instance the case for the Whistler-Helicon modes discussed above since $\beta$ depends on plasma parameters, one of which being the electron gyro-frequency. The in-depth study of these challenges, and from there a critical assessment of the potential of image rotation for plasma rotation diagnostics, is is left for future studies.

\section{Summary}
\label{Sec:summary}

Wave propagation properties in a moving medium differ from those in this same medium at rest. In the case of a rotating motion, different manifestations have been known to occur in isotropic dielectrics.  In this study we discuss some recents findings obtained when considering a magnetised plasma.

First, it is shown that the polarisation drag effect known to occur in rotating isotropic dielectrics is recovered in the case of a rotating magnetised plasma. Magnetised plasmas do however bring more complexity in that this polarisation drag is now superimposed onto the classical Faraday rotation which arises from the magnetic field. Polarisation rotation in a rotating magnetised plasma where the wave vector is aligned with both the rotation axis and the magnetic field thus has two contributions: one from the magnetic field and one from rotation. The latter had been neglected in plasmas up to this date.

While this new contribution is often negligible compared to Faraday rotation, we showed that polarisation drag can be the dominant contribution to polarisation rotation under particular conditions. This is notably true for underdense plasmas in a frequency band below the ion cyclotron frequency where polarisation drag is enhanced. Exploring this further, we showed that this effect - if controlled - could lead to unique nonreciprocal properties at THz frequencies, opening opportunities for the development of compact wavelength-agile isolators using rotating magnetised plasmas. Besides laboratory experiments, we also showed that polarisation drag in plasmas may also be of importance in astrophysics and more particularly for pulsar physics. More specifically, pulsar polarimetry is among other things routinely used to infer galactic magnetic fields, and it is shown that failing to account for this new effect in the magnetosphere that is generally assumed to be co-rotating with the pulsar could lead to error in magnetic field estimates. Fortunately, we showed that the different signature of polarisation drag at low observation frequency could enable disambiguating this effect in pulsars' signal.

Second, it is demonstrated that the phenomenon of image rotation, that is the rotation of the transverse structure of a wave in the presence of rotation which had been known to occur in isotropic dielectrics, is also found for a number of low frequency magnetised plasma waves in rotating plasmas. Just like polarisation drag arises from a phase shift induced by rotation between eigenmodes with opposite spin angular momentum, image rotation is also shown to result from a phase shift induced by rotation but this time between eigenmodes with opposite orbital angular momentum. Although the implications of this effect of rotation on the wave's orbital angular momentum are yet to be uncovered, the fact that image rotation results only from rotation and not from the magnetic field itself creates opportunities for the development of new plasma rotation diagnostic tools.

To conclude, the interpretation of polarisation drag and image rotation in plasmas as two different coupling mechanisms between the wave's spin and orbital angular momentum components and medium's rotation proposed here opens a number of new research directions. A particularly interesting prospect will be to extend this work considering only the reactive response of a rotating plasma to now include the active (or resonant) rotating plasma response. Understanding active coupling will indeed help advance our comprehension of the practicality of using waves to control plasma rotation~\cite{Fisch1992,Fetterman2008,Ochs2021,Ochs2021a,Ochs2022}, which holds promise both for alternative fusion concepts~\cite{Rax2017,Ochs2017b} and for plasma mass separation applications~\cite{Gueroult2018,Zweben2018}. It will also provide the tools to study how waves can generate magnetic fields~\cite{Ochs2020} when rotation is now present, possibly shedding light onto the complex interplay between rotation and magnetic field dynamics observed across many environments in astrophysics~\cite{Kulsrud1999,Miesch2009}.

\section*{Acknowledgements}

RG acknowledges the support of the French Agence Nationale de la Recherche (ANR), under grant ANR-21-CE30-0002 (project WaRP). The authors would also like to thank Julien Langlois for constructive discussions.

\section*{References}
\bibliography{References.bib,EPS2022.bib}
\bibliographystyle{unsrt}

\end{document}